\documentclass[iop]{emulateapj}
\usepackage{apjfonts}

\newcommand{\f}{\phantom{2}}

\newcommand{\ltsimeq}{\raisebox{-0.6ex}{$\,\stackrel 
        {\raisebox{-.2ex}{$\textstyle <$}}{\sim}\,$}} 
\newcommand{\gtsimeq}{\raisebox{-0.6ex}{$\,\stackrel 
        {\raisebox{-.2ex}{$\textstyle >$}}{\sim}\,$}}

\newcommand{\mgii}{Mg\,{\sc ii}}

\newcommand{\lya}{Ly\,$\alpha$}
\newcommand{\lyb}{Ly\,$\beta$}
\newcommand{\lyg}{Ly\,$\gamma$}
\newcommand{\nv}{N\,{\sc v}}

\newcommand{\oii}{[O\,{\sc ii}]}

\newcommand{\myemail}{chris.willott@nrc.ca}

\def\hst{{\it Hubble Space Telescope~}}
\def\co21{CO\,(2-1)}

\shorttitle{The CFHQS: nine new quasars and the luminosity function at z=6}
\shortauthors{Willott et al.}

\begin{document}


\title{The Canada-France High-$z$ Quasar Survey: \\nine new quasars and the luminosity function at redshift 6}


\author{
Chris J. Willott\altaffilmark{1},
Philippe Delorme\altaffilmark{2},
C\'eline Reyl\'e\altaffilmark{3},
Loic Albert\altaffilmark{4},
Jacqueline Bergeron\altaffilmark{5},
David Crampton\altaffilmark{1},
Xavier Delfosse\altaffilmark{6},
Thierry Forveille\altaffilmark{6},
John B. Hutchings\altaffilmark{1},
Ross J. McLure\altaffilmark{7},
Alain Omont\altaffilmark{5},
and David Schade\altaffilmark{1},
}

\altaffiltext{1}{Herzberg Institute of Astrophysics, National Research Council, 5071 West Saanich Rd, Victoria, BC V9E 2E7, Canada; \myemail}
\altaffiltext{2}{School of Physics and Astronomy, University of St Andrews, North Haugh, St Andrews, KY16 9SS, UK}
\altaffiltext{3}{Observatoire de Besan\c{c}on, Universit\'e de Franche-Comt\'e, Institut Utinam, UMR CNRS 6213, BP1615, 25010 Besan\c{c}on Cedex, France}
\altaffiltext{4}{Canada-France-Hawaii Telescope Corporation, 65-1238 Mamalahoa Highway, Kamuela, HI96743, USA}
\altaffiltext{5}{Institut d'Astrophysique de Paris, CNRS and Universit\'e Pierre et Marie Curie, 98bis Boulevard Arago, F-75014, Paris, France}
\altaffiltext{6}{Laboratoire d'Astrophysique, Observatoire de Grenoble, Universit\'e J. Fourier, BP 53, F-38041 Grenoble, Cedex 9, France}
\altaffiltext{7}{Scottish Universities Physics Alliance, Institute for Astronomy, University of Edinburgh, Royal Observatory, Blackford Hill, Edinburgh, EH9 3HJ, UK}

\begin{abstract}

We present discovery imaging and spectroscopy for nine new $z \sim 6$
quasars found in the Canada-France High-$z$ Quasar Survey (CFHQS)
bringing the total number of CFHQS quasars to 19. By combining the
CFHQS with the more luminous SDSS sample we are able to derive the
quasar luminosity function from a sample of 40 quasars at redshifts
$5.74<z<6.42$.  Our binned luminosity function shows a slightly lower
normalisation and flatter slope than found in previous work. The
binned data also suggest a break in the luminosity function at
$M_{1450} \approx -25$. A double power law maximum likelihood fit to
the data is consistent with the binned results. The luminosity
function is strongly constrained (1\,$\sigma$ uncertainty $<0.1$\,dex)
over the range $-27.5<M_{1450}<-24.7$. The best-fit parameters are
$\Phi(M_{1450}^{*})= 1.14\times 10^{-8}\,{\rm Mpc}^{-3}\,{\rm
  mag}^{-1}$, break magnitude $M_{1450}^{*}=-25.13$ and bright end
slope $\beta=-2.81$. However the covariance between $\beta$ and
$M_{1450}^{*}$ prevents strong constraints being placed on either
parameter. For a break magnitude in the range $-26 < M_{1450}^{*} <
-24$ we find $-3.8 < \beta < -2.3$ at 95\% confidence. We calculate
the $z=6$ quasar intergalactic ionizing flux and show it is between 20
and 100 times lower than that necessary for reionization. Finally, we
use the luminosity function to predict how many higher redshift
quasars may be discovered in future near-IR imaging surveys.

\end{abstract}


\section{Introduction}

Observations of the most distant quasars provide important information
on the state of the universe within the first billion years. They are
used to probe the transition from a mostly neutral to mostly ionized
intergalactic medium (IGM). They provide the only direct estimates of
the metallicities and evolutionary states of massive galaxies. Quasars
also probe the very early growth of supermassive black holes in the
centers of massive galaxies.

The success of surveys such as the Sloan Digital Sky Survey (SDSS; Fan
et al. 2006a; Jiang et al. 2009) and Canada-France High-$z$ Quasar
Survey (CFHQS; Willott et al. 2009) means that the number of known
quasars at $z\sim 6$ is approaching 50. These samples are now large
enough to determine the typical properties of quasars over a range of
luminosities. One of the most remarkable results is that in most
respects, quasars at $z\sim 6$ have properties such as metallicity,
emission line strength, hot dust mid-IR luminosity and cool dust
far-IR luminosity very similar to lower redshift quasars
(Freudling et al. 2003; Iwamuro et al. 2004; Fan et al. 2004; Jiang et
al. 2006; Wang et al. 2008).

The number of known $z\sim 6$ quasars is now large enough to allow the
luminosity function to be determined. The luminosity function is
important since it encodes information about the build-up of
supermassive black holes and gives the hard ionizing radiation output
of the quasar population. Early work on the $z\sim 6$ quasar
luminosity function was based on determining the normalisation,
$\Phi$, and bright-end power law slope, $\beta$, from the SDSS. This
was done by direct determination from the known SDSS quasars by Fan et
al. (2004) yielding $\Phi$ a factor of 30 times lower than at $z=3$
and $\beta=-3.2$ (95\% confidence range of $-4.2$ to $-2.2$).  Further
constraints came from the lack of gravitational lenses amongst the
known SDSS quasars which implied $\beta \gtsimeq -4$ (Fan et al. 2003;
Richards et al. 2004). Jiang et al. (2008; 2009) combined the SDSS
main sample with the SDSS deep stripe to find a somewhat shallower
slope of $\beta=-2.6 \pm 0.3$. However, the binned data shown in Jiang
et al. (2009) suggest that there may be a break in the luminosity
function at $M_{1450} \approx -26$, in which case $\beta$ could be
closer to the value determined by Fan et al. (2004).

There have been several small area surveys searching for faint $z\sim
6$ quasars. These are summarized by Shankar \& Mathur (2007). Only the
key results for surveys at $z>5.7$ are repeated here. Cool et
al. (2006) discovered three quasars at $z>5$, including one quasar at
$z=5.85$ with $z'=20.68$, in the AGES survey which covers only 8.5 sq
deg. As shown by Jiang et al. (2008), Cool et al. were very fortuitous
to discover such a bright quasar at $z>5.7$ in such a small sky
area. The space density of similar luminosity quasars determined by
Jiang et al. from the much larger SDSS deep stripe is about six times
lower than that of Cool et al. 

Mahabal et al. (2005) searched for high-redshift quasars around known
SDSS quasars and serendipitously discovered one with magnitude
$z'=23.0$ at $z=5.70$ foreground to a SDSS quasar located at
$z=6.42$. In total, Mahabal et al. searched 2.5 sq deg. to a depth of
$z'=24.5$. Willott et al. (2005a) searched the first data release
(T0001) of the CFHT Legacy Survey Deep and found no quasars to
$z'=23.35$ in 3.8 sq. deg. The Mahabal et al. and Willott et
al. results reveal that there must be a break in the luminosity
function between these low luminosities ($M_{1450} \approx -23$) and
the high luminosity SDSS quasars ($M_{1450} \approx -27$). Because low
luminosity quasars provide the bulk of ionizing photons from AGN,
Willott et al. (2005a) showed that quasars were incapable of
reionizing the IGM at $z=6$, consistent with constraints from the
unresolved X-ray background (Djikstra et al. 2004).

\begin{figure}
\hspace{0.7cm}
\resizebox{0.4\textwidth}{!}{\includegraphics{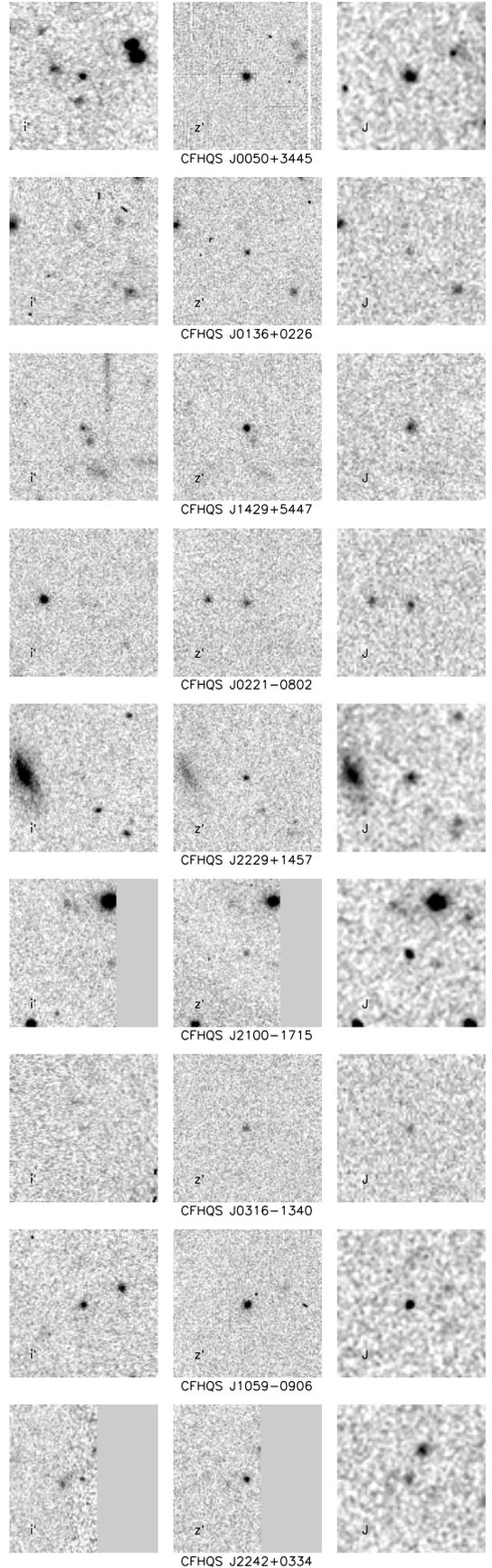}}
\caption{Images in the $i'$, $z'$ and $J$ filters centered on the nine
  CFHQS quasars. Each image covers $20'' \times 20''$. The images are
  oriented with north up and east to the left. Two quasars are located close to the edges of the chips at $i'$ and $z'$ and therefore do not have data across the whole region.
\label{fig:cutouts}
}
\end{figure}

\begin{figure*}
\resizebox{0.93\textwidth}{!}{\includegraphics{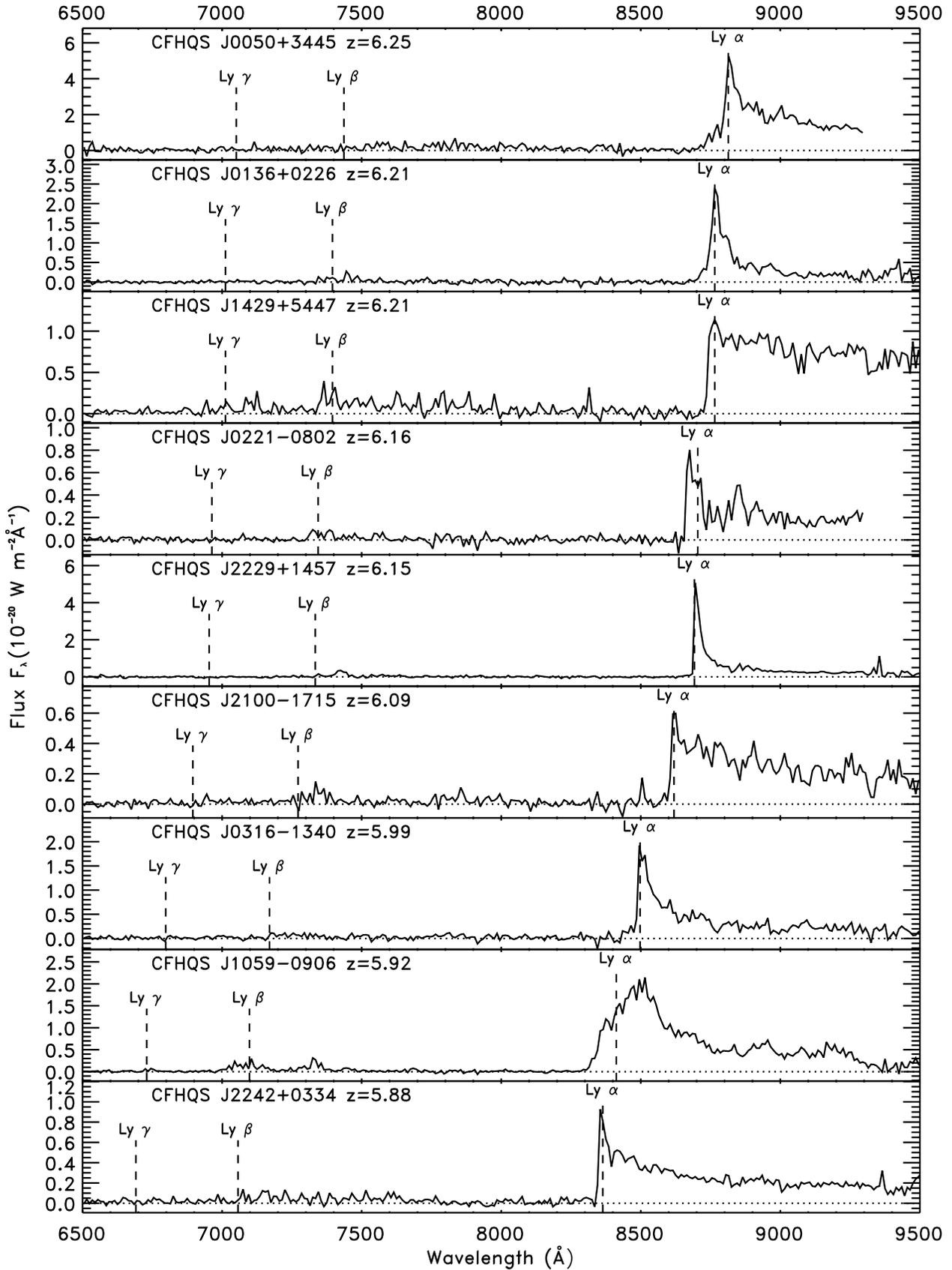}}
\caption{Optical spectra of the nine newly discovered quasars. The
expected locations of \lya, \lyb\ and \lyg\ are marked with dashed
lines.  All spectra are binned in 10\,\AA\ pixels. The dashed
\lya\ lines often do not appear to line up with the peak flux in
these plots (even though most redshifts are measured from the
\lya\ peak), due to a combination of the binning and asymmetric line
profiles.
\label{fig:spec}
}
\end{figure*}

In this paper we present the discovery of nine further CFHQS quasars
and use a combined CFHQS/SDSS sample to derive the $z=6$ quasar
luminosity function. Sec.\,2 presents the data on these quasars and
notes on each one are given in Sec.\,3. Sec.\,4 discusses the depth,
area and completeness of the parts of the CFHQS and SDSS which are
used in this paper to determine the luminosity function. Sec.\,5
presents binned and parametric derivations of the $z=6$ quasar
luminosity function. Sec.\,6 discusses what this new derivation
implies for the quasar intergalactic ionizing flux and search
strategies for even higher redshift quasars. We present the
conclusions in Sec.\,7.

All optical and near-IR magnitudes in this paper are on the AB
system. Cosmological parameters of $H_0=70~ {\rm km~s^{-1}~Mpc^{-1}}$,
$\Omega_{\mathrm M}=0.28$ and $\Omega_\Lambda=0.72$ (Komatsu et
al. 2009) are assumed throughout.


\begin{deluxetable*}{c l c c c c c}
\tablewidth{500pt}
\tablecolumns{8}
\vspace{0.4cm}
\tablecaption{\label{tab:photom} Quasar positions and photometry} 
\tablehead{ Quasar    & RA and DEC (J2000.0)     &  $i'$ mag        &    $z'$ mag       &      $J$ mag      &  $i'-z'$        &  $z'-J$     }
\startdata
CFHQS\,J005006+344522 &  00:50:06.67 +34:45:22.6 & $23.49 \pm 0.04$ &  $20.47 \pm 0.03$ & $19.89 \pm 0.04$  & $3.02 \pm 0.05$ & $0.58 \pm 0.05$  \\
CFHQS\,J013603+022605 &  01:36:03.17 +02:26:05.7 & $>24.72^{\rm a}$  &  $22.10 \pm 0.09$ & $22.09 \pm 0.22$  & $>2.62$         & $0.01 \pm 0.24$  \\
CFHQS\,J142952+544717 &  14:29:52.17 +54:47:17.7 & $23.88 \pm 0.06$ &  $21.45 \pm 0.03$ & $20.64 \pm 0.07$  & $2.43 \pm 0.07$ & $0.81 \pm 0.08$  \\
CFHQS\,J022122-080251 &  02:21:22.71 -08:02:51.5 & $25.60 \pm 0.21$ &  $22.63 \pm 0.05$ & $22.03 \pm 0.14$  & $2.97 \pm 0.22$ & $0.60 \pm 0.15$  \\
CFHQS\,J222901+145709 &  22:29:01.65 +14:57:09.0 & $24.89 \pm 0.16$ &  $22.03 \pm 0.05$ & $21.95 \pm 0.07$  & $2.86 \pm 0.17$ & $0.08 \pm 0.09$  \\
CFHQS\,J210054-171522 &  21:00:54.62 -17:15:22.5 & $24.55 \pm 0.21$ &  $22.35 \pm 0.09$ & $21.42 \pm 0.10$  & $2.20 \pm 0.23$ & $0.93 \pm 0.13$  \\
CFHQS\,J031649-134032 &  03:16:49.87 -13:40:32.3 & $24.57 \pm 0.19$ &  $21.72 \pm 0.08$ & $21.77 \pm 0.17$  & $2.85 \pm 0.21$ & $-0.05 \pm 0.19$ \\
CFHQS\,J105928-090620 &  10:59:28.61 -09:06:20.4 & $23.21 \pm 0.04$ &  $20.82 \pm 0.03$ & $20.79 \pm 0.07$  & $2.39 \pm 0.05$ & $0.03 \pm 0.08$  \\
CFHQS\,J224237+033421 &  22:42:37.55 +03:34:21.6 & $24.22 \pm 0.12$ &  $21.93 \pm 0.04$ & $22.13 \pm 0.12$  & $2.29 \pm 0.13$ & $-0.20 \pm 0.13$ \\
\enddata
\tablecomments{All magnitudes are on the AB system.}
\tablenotetext{a}{Where not detected at $> 2\sigma$ significance, a $2\sigma$ lower limit is given.}
\end{deluxetable*}

\begin{deluxetable*}{c c c c c c c c}
\tablewidth{500pt}
\tablecolumns{8}
\vspace{0.4cm}
\tablecaption{\label{tab:specobs} Optical spectroscopy observations of the new CFHQS quasars}
\tablehead{ Quasar & Redshift & Date~~~~~~~ & Resolving & Slit Width & Exp. Time & Seeing   & $M_{1450}$ \\
                   & $z$      &             &  Power    &  (Arcsec)  &   (s)     & (Arcsec) &           }
\startdata
CFHQS\,J0050+3445 & 6.25 &  2008 Sep 25                & 4800 & 1.0 & \f 6250 & 1.3 & $-26.62$  \\
CFHQS\,J0136+0226 & 6.21 &  2009 Aug 02 + 2009 Sep 13  & 1300 & 1.0 & \f 5400 & 0.9 & $-24.40$  \\
CFHQS\,J1429+5447 & 6.21 &  2009 Mar 31 + 2009 Apr 30  & 1300 & 1.0 & \f 5400 & 0.9 & $-25.85$  \\
CFHQS\,J0221-0802 & 6.16 &  2008 Oct 24                & 1300 & 1.0 & \f 5400 & 0.7 & $-24.45$  \\
CFHQS\,J2229+1457 & 6.15 &  2008 Sep 24                & 4800 & 1.0 & \f 3600 & 0.9 & $-24.52$  \\
CFHQS\,J2100-1715 & 6.09 &  2008 Jul 30                & 1300 & 1.0 & \f 3600 & 0.7 & $-25.03$  \\
CFHQS\,J0316-1340 & 5.99 &  2009 Aug 02                & 1300 & 1.0 & \f 3600 & 0.7 & $-24.63$  \\
CFHQS\,J1059-0906 & 5.92 &  2009 Mar 22                & 1300 & 1.0 & \f 3600 & 0.6 & $-25.58$  \\
CFHQS\,J2242+0334 & 5.88 &  2008 Oct 18                & 1300 & 1.0 & \f 3600 & 0.5 & $-24.22$  \\
\enddata
\tablecomments{GMOS spectra have resolving power 1300 and ESI spectra have resolving power 4800. As in Willott et al. (2009), absolute magnitudes ($M_{1450}$) are calculated using the measured $J$-band magnitudes and assuming a template quasar spectrum. Note that this is slightly different to the method used in Willott et al. (2007), which was based on measuring the continuum redward of \lya\ in the spectrum and assuming a spectral index.}
\end{deluxetable*}

\section{New CFHQS quasars}

The optical imaging phase of the CFHQS and associated Canada-France
Brown Dwarf Survey (CFBDS) is now complete. It draws on the Very Wide
(VW), Wide and Deep components of the CFHT Legacy Survey (CFHTLS), the
Subaru/XMM-Newton Deep Survey (SXDS) and the Red-sequence Cluster
Survey (RCS-2). For some of these surveys, additional imaging was
obtained in PI mode to complete the necessary filter coverage. Further
details of the observations are given in Willott et
al. (2009). Near-IR and spectroscopic follow-up is still
ongoing, so this paper does not present the final results of the
CFHQS. Sec.\,\ref{surveys} discusses the complete parts of the CFHQS
which are used to determine the quasar luminosity function in this
paper.

Quasar candidates were selected using the same color criteria as in
Willott et al. (2009). The first cut is at $i'-z'>2$ to select very
red objects which are a mixture of brown dwarfs and high-redshift
quasars. Pointed $J$-band photometry with the ESO New Technology
Telescope and Nordic Optical Telescope is then used to separate these
two types of object, the dwarfs having redder $z'-J$ colors than the
quasars. Spectroscopy of candidates was carried out mostly using the
GMOS spectrographs at the twin Gemini telescopes. Two quasars
(CFHQS\,J0050+3445 and CFHQS\,J2229+1457) were observed with the ESI
spectrograph at Keck-II. Spectroscopic reductions were performed as
detailed in Willott et al. (2007).

Table\,\ref{tab:photom} gives the positions and photometry for the
nine new quasars. Table\,\ref{tab:specobs} details the spectroscopic
observations. Note in this table and the rest of the paper, quasar
names use an abbreviated form. Fig.\,\ref{fig:cutouts} shows cutout
images centred on each quasar and Fig.\,\ref{fig:spec} plots their
optical spectra. Larger area finding charts are in the Appendix.

\section{Notes on individual quasars}
\label{indiv}

\subsection{CFHQS\,J0050+3445}
The redshift of the quasar is $z=6.25$ based on the narrow peak of
\lya\ emission on top of the broad line. This is consistent with the
redshift based on the broad \mgii\ emission line (Willott et al. in
prep.).  This is the second highest redshift in the CFHQS. It is also
the second most luminous CFHQS quasar with $M_{1450}=-26.62$, which is
nearly as luminous as some of the SDSS quasars of Fan et al. (2006b).

This quasar is included in the group of RCS-2/CFHTLS VW quasars,
however it is the only one in that group which is not actually located
in these survey areas. We carried out $z'$ band imaging of
regions already observed at $i'$ band with CFHT of the intragroup
environment of the Local Group (PI R. Ibata). These observations were
necessary to use scheduled time (supposed to be $i'$ band followup of
RCS-2 fields) when the $i'$ filter broke and was unusable. This field
proved not to be ideal for the CFHQS because of several Local Group
variable red giant stars which had non-contemporaneous $i'z'J$
photometry appearing to be $z\sim 6$ quasars, which required repeat
photometry to eliminate.

\subsection{CFHQS\,J0136+0226}
The spectrum shows a fairly narrow broad \lya\ line. The redshift of
$z=6.21$ is derived from the peak of the narrow \lya. This is the only
quasar in this paper which was not detected in the $i'$ filter. It has
a lower limit on the color of $i'-z'>2.62$.  The $z'-J=0.01$ color is
quite blue for such a high redshift quasar. This is likely because of
the high equivalent width \lya\ emission.

The faint galaxy 3\,arcsec north of the quasar
(Fig.\,\ref{fig:cutouts}) happened to be located along the
spectrograph slit. It has a single emission line at 7468\AA\ which is
likely \oii\ at $z=1.00$. The low luminosity of this galaxy (at
observed frame wavelengths from $i'$ to $J$ band) indicates it has
only a moderate mass and at this separation from the quasar will
provide negligible gravitational lensing magnification (see
e.g. Sec.\,4 of Willott et al. 2005b).

\subsection{CFHQS\,J1429+5447}
This quasar was discovered in the CFHTLS Wide W3 region. However, it
is much brighter than our Wide magnitude limit of $z'=23$ and has the
second brightest absolute magnitude in this paper. The spectrum shows
a strong continuum with only a weak \lya\ emission line at $z=6.21$. This is
consistent with the relatively red color of $z'-J=0.81$. The spectrum
shows a dark \lya\ absorption trough from $5.85<z<6.16$ with lots of
sharp \lya\ peaks at lower redshift. A more sensitive and higher
spectral resolution spectrum is required for a complete analysis.

There is an obvious \mgii\ absorption doublet at 9061.5,
9083.8\,\AA\ which corresponds to $z=2.241$. The apparent broad
absorption at 9300\,\AA\ is telluric absorption which has not been
corrected for (and can be seen for several other quasars in
Fig.\,\ref{fig:spec}).

\subsection{CFHQS\,J0221-0802}

Another quasar from the CFHTLS Wide (W1 field) and the faintest at
$z'$ band in this paper.  The \lya\ emission peaks at $z=6.13$ with a
narrow spike atop an asymmetric broad component. There is also a
likely \nv\ $\lambda 1240$ line consistent with the \lya\ redshift. On
the other hand, the near-IR spectrum (Willott et al. in prep.)
shows the \mgii\ line at $z=6.16$ which we adopt as the systemic redshift for
this quasar.

\subsection{CFHQS\,J2229+1457}

This is another CFHQS quasar with a strong, very narrow \lya\ emission
line. The best-fit redshift to the peak of the \lya\ line gives
$z=6.15$. An unpublished near-IR spectrum shows the \mgii\ line to
have an identical redshift.

\subsection{CFHQS\,J2100-1715}

This quasar has a very similar spectrum to CFHQS\,J0221-0802 with a
narrow \lya\ peak at $z=6.09$ (consistent with the unpublished \mgii\ line
redshift). There is an interesting peak in the \lya\ forest at
$z=6.00$ which appears too far from the systemic redshift to be within
the quasar-ionized near-zone for a quasar of such moderate
luminosity. CFHQS\,J2100-1715 has a rather red color of $z'-J=0.93$
for its redshift, which may be partly due to the weak \lya\ line, but
also indicates a red spectral slope.

\subsection{CFHQS\,J0316-1340}
The spectrum shows a double-peaked \lya\ emission line. We take the
redshift of $z=5.99$ from the blueward peak. In terms of the spectrum
and photometry, there is nothing unusual about this quasar.

\subsection{CFHQS\,J1059-0906}
This is one of the few CFHQS quasars to have a very broad \lya\ line,
typical of those in the SDSS main sample. The quasar is fairly bright
($z'=20.82$), consistent with the idea that line width correlates with
luminosity due to the black hole mass. There is no narrow peak to
\lya\ and consequently a large uncertainty on the redshift which is
estimated to be $z=5.92$ based on the full broad \lya\ profile and
expected asymmetry due to hydrogen absorption.

\subsection{CFHQS\,J2242+0334}
Yet another quasar with a narrow \lya\ peak just redward of a sharp
cutoff. At $z=5.88$ and $M_{1450}=-24.22$ this is the lowest redshift
and least luminous quasar from the RCS-2 and CFHTLS Very Wide. It is
also the bluest quasar in the CFHQS with $z'-J=-0.2$. This is not too
surprising because the expected $z'-J$ colors of $5.5<z<6.7$ quasars
have a minimum at $z=5.8$ (Willott et al. 2009).

\begin{figure*}
\resizebox{0.96\textwidth}{!}{\includegraphics{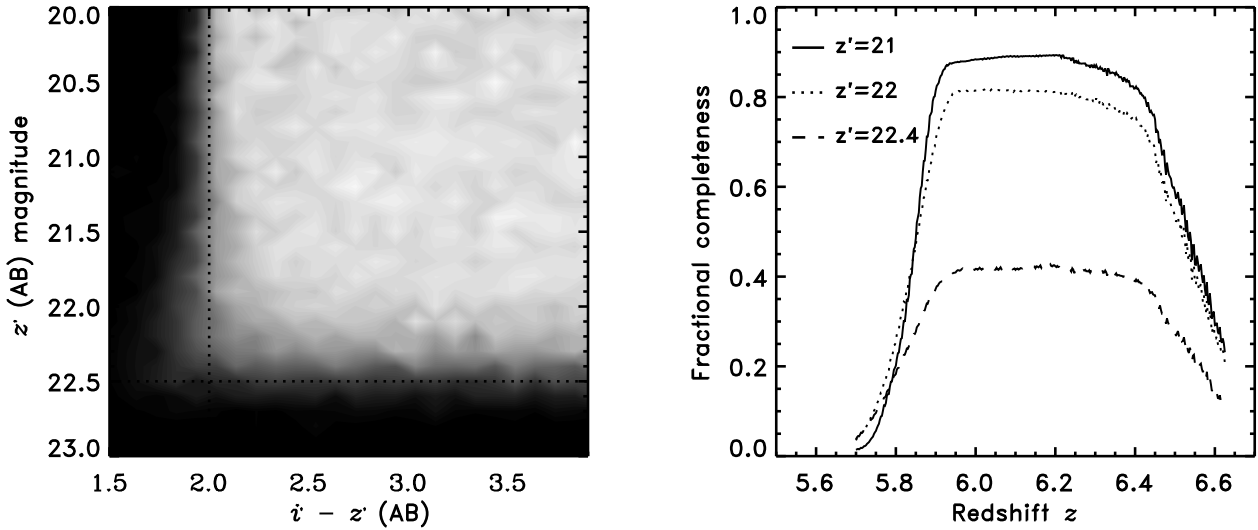}}
\caption{Left: Photometric completeness as a function of artificial source magnitude and color for a typical RCS-2 patch, CDE2329. White corresponds to fractional completeness of 1 and black is 0. The dotted lines show the selection criteria $z'<22.5$ and $i-z>2$. Note that a small fraction of objects with true magnitudes and colors outside of the selection criteria are selected due to photometric errors. Right: Completeness as a function of redshift for the same patch using the set of cloned lower redshift quasars. Lines are plotted at three different apparent magnitude levels. 
\label{fig:photcomp}
}
\end{figure*}

\section{CFHQS and SDSS $z=6$ quasar surveys}
\label{surveys}

\subsection{CFHQS}

The sample of CFHQS quasars used to derive the luminosity function in
this paper contains all the quasars discovered from patches
(contiguous sub-regions) which have complete near-IR imaging and
spectroscopic follow-up to the full magnitude and color limits. The
CFHTLS Wide areas are not included because none of the Wide patches
are yet complete. All the quasars from the CFHTLS Deep/SXDS, RCS-2 and
CFHTLS VW are located in complete patches. Therefore the CFHQS sample
in this paper consists of 12 quasars from the RCS-2 (including
CFHQS\,J0050+3445), 4 quasars from the CFHTLS VW and 1 quasar from the
CFHTLS Deep/SXDS. The effective sky areas are 494 sq. deg. for the
combination of RCS-2 and CFHTLS VW and 4.47 sq. deg. for the CFHTLS
Deep/SXDS. We use the fifth data release (T0005) of the CFHTLS Deep. One of the quasars included in the CFHQS RCS-2 sample was
actually discovered by the SDSS (SDSS\, J2315-0023 at $z=6.12$; Jiang
et al. 2008), but as explained in Willott et al. (2009) it was already
in our follow-up list before we learned of its discovery. Since this
quasar lies within the CFHQS area, it must be included in our sample
if we are not to underestimate the number of quasars.

The patches have variable optical imaging depth and therefore careful
completeness corrections have to be applied. The method to determine
completeness consisted of inserting 1.5 million artificial point
sources into our data and determining the fraction recovered by our
selection pipeline as a function of magnitude and color. Extensive
details of this process are given in the CFBDS paper of Reyl\'e et
al. (2009). An example of the resulting completeness as a function of
apparent magnitude and color is shown in the left hand panel of
Fig.\,\ref{fig:photcomp}. 

For the CFHQS quasars, rather than the brown dwarfs in the CFBDS,
several further steps are necessary. Galactic extinction affects both
the observed color and absolute magnitude, so is corrected for based
on the maps of Schlegel et al. (1998). For quasars we need to
determine how the color selection cut and completeness affects the
redshift completeness. This is particularly important at $z\approx
5.8$ where the typical quasar color is approximately the same as our
color lower limit. We use the 180 cloned $z\approx 3.1$ SDSS quasars
first discussed in Willott et al. (2005a) to determine how the
completeness as a function of color translates to completeness as a
function of redshift. This process assumes that the spectra of quasars
do not evolve significantly from $z=3$ to $z=6$ in accord with most
observations (Fan et al. 2004). The only factor that evolves with
redshift is the transparency of the IGM to \lya\ radiation. This is
modeled using the empirical fit of Songaila (2004). The
proximity of the CFHQS quasars in color-color space to their expected
locations (Willott et al. 2009) shows that the cloning process is very
effective.

The right hand panel of Fig.\,\ref{fig:photcomp} shows the
completeness as a function of redshift for a typical patch from the
RCS-2. Curves are plotted for three different apparent
magnitudes. This shows that the completeness does not change much from
$z'=21'$ to $z'=22$. By the time the magnitude limit is reached, the
completeness for this RCS-2 patch is about half of its peak
value. Part of this decline is due to some regions having 10\,$\sigma$
$z'$ limits brighter than $z'=22.5$ and part is due to photometric
scatter (the hard cut at $z'=22.5$ misses some objects with intrinsic
magnitudes of $z'=22.4$).  The $i'-z'>2$ cut selects quasars at
$z\gtsimeq 5.8$.  Most of the lowest redshift CFHQS quasars are
expected to be close to the magnitude limit, because the higher
magnitude errors increase the number of objects with true colors
$i'-z'<2$ appearing in the selection box. At $z \gtsimeq 6.5$ quasars
move out of the selection box because their $z'-J$ colors become like
those of late T dwarfs and the absolute magnitude at a fixed $z'$
magnitude decreases rapidly.

\begin{figure}
\resizebox{0.50\textwidth}{!}{\includegraphics{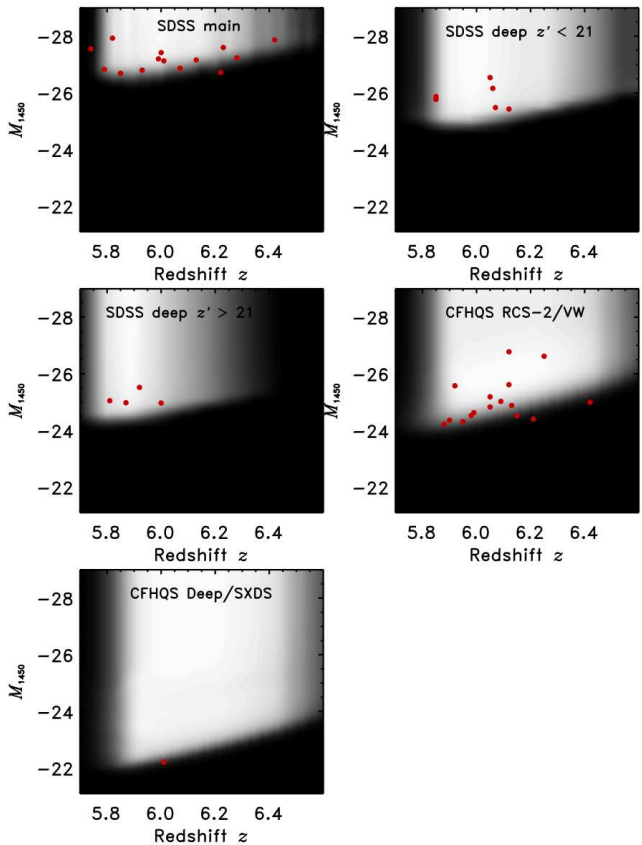}}
\caption{Completeness as a function of absolute magnitude and redshift for all SDSS and CFHQS samples used in this paper. White equals peak completeness and black equals zero. The red circles show the quasars discovered by each sample. SDSS deep $z'<21$ corresponds to Jiang et al. (2008) and SDSS deep $z'>21$ to Jiang et al. (2009).  
\label{fig:comparrays}
}
\end{figure}

The final step is to convert the CFHQS completeness as functions of
redshift and apparent magnitude into functions of redshift and
absolute magnitude. This is accomplished by using the 180 cloned lower
redshift quasars to mimic the expected variance in the relationship
between apparent and absolute magnitude at a fixed
redshift. Fig.\,\ref{fig:comparrays} shows two panels of the
completeness as a function of redshift and absolute magnitude for
CFHQS. One panel is for all the CFHQS RCS-2/VW patches and one is for
the combination of the CFHTLS Deep and SXDS. The locations of quasars
in the complete samples used in this paper are also marked. Most
quasars were found close to the edge of the selection function in
absolute magnitude. The two quasars at $z \ge 6.2$ in very dark
regions of the plot are in regions where the completeness is only
$\sim 5\%$. The completeness is so low here because only parts of some
patches have high completeness to the magnitude limit of
$z'=22.5$. Effectively, this is equivalent to only a fraction of the full sky
area having been surveyed to the full depth.

\subsection{SDSS main sample}

The bright end of the $z=6$ quasar luminosity function is best sampled
by the SDSS main sample described in Fan et al. (2000; 2001; 2003;
2004; 2006b). The most recent published sample from Fan et al. (2006b)
contains 14 quasars at $5.74<z<6.42$ with uniform selection criteria
found from a total sky area of $\sim 6600$ sq. deg. The quasars are
selected to a magnitude limit of $z'<20.2$ and color cut of
$i'-z'>2.2$. Note that due to differences between the filters used by
SDSS and CFHT, the SDSS sample has better completeness at $z\approx
5.8$ than the CFHQS, despite the apparently higher color cut of the
SDSS. This is the reason why the CFHQS sample has a slightly higher
median redshift ($z=6.05$) than the SDSS sample ($z=6.0$).

The SDSS main sample completeness as a function of absolute magnitude
and redshift was supplied by X.\,Fan and is that shown in Fig.\,7 of
Fan et al. (2003). Fan et al. use a slightly different cosmological
model to us and therefore their selection function and quasar absolute
magnitudes are both converted to our cosmology.

The SDSS main sample completeness as a function of absolute magnitude
and redshift is also shown in Fig.\,\ref{fig:comparrays}. The SDSS
quasars cluster very strongly to the edge of the selection function
indicating a fairly steep bright end slope to the luminosity
function. Indeed, Fan et al. (2004) derived a slope of $\beta=-3.2$
from these data. Two SDSS main quasars are found in regions of very
low selection probability ($\sim 5\%$), as was also found for the CFHQS.

\subsection{SDSS deep stripe}

A small portion of the SDSS, the deep stripe, has repeated imaging
which allows one to search for fainter quasars than can be found in
the main sample. Jiang et al. (2008) constructed a sample of six
quasars at $20<z'<21$ in a sky area of 260 sq. deg. In a following
paper they extended their search to fainter fluxes and found a
complete sample of four quasars at $21<z'<21.8$ in 195 sq. deg (Jiang
et al. 2009). These 10 quasars have redshifts between $5.78$ and
$6.12$.

The completeness for the SDSS deep stripe has been determined by Jiang
et al. (2008; 2009). In their first paper, quasars were selected using
follow-up in either the $J$ or $H$ filters (the $J$ filter is optimum,
but was unavailable for some observations). Therefore there are two
different selection functions for the $J$ and $H$ samples (Fig.\,5 of
Jiang et al. 2008). We combine the $J$ and $H$ selection functions in
proportion to the fraction of quasars discovered with each
filter. Note that there are only small differences between these two
functions and mostly at the high redshift end where no quasars were
detected. The deeper sample of Jiang et al. (2009) covers some of the same
area as their brighter sample. We therefore used the brighter sample selection function
for 65 sq. deg. and for the remaining 195 sq deg. set the selection
function to be the greater of either the bright or faint sample
functions. This was necessary because at some locations in the
$M_{1450},z$ plane, the brighter selection function is actually higher
than the fainter selection function.

Fig.\,\ref{fig:comparrays} shows two panels for the completeness of
the two SDSS deep stripe samples. The SDSS deep $z'>21$ sample reaches
to within half a magnitude of the CFHQS RCS-2/VW sample, but covers
less than half the area on the sky. It is apparent that quasars from
both the SDSS deep samples do not cluster as strongly to the selection
function as either the SDSS main or CFHQS quasars.

As mentioned above, SDSS\, J2315-0023 from Jiang et al. (2008) was
also discovered in CFHQS imaging and is the only quasar to have
been identified in the small fraction of overlap in sky area between
the CFHQS and SDSS deep stripe. This quasar will be included in both
the CFHQS and SDSS deep stripe samples for the purposes of calculating
the luminosity function and the sky areas of the samples will be
treated as if they are independent. Excluding this duplication, the
combined CFHQS and SDSS $z=6$ quasar sample used in this paper
consists of 40 quasars at redshifts $5.74<z<6.42$.

\section{z=6 quasar luminosity function}

\subsection{Binned luminosity function}

To help determine the appropriate parametric form of the luminosity
function we first carry out a simple binning of the data. As is well
known, the main disadvantage of this method is that for small samples
with inhomogeneous coverage of parameter space, the choice of bins can
make large differences to the results. In addition, binned luminosity
functions are difficult to extrapolate beyond well-sampled parts of
parameter space.

We use the binned $1/V_{a}$ method of Avni \& Bahcall (1980) where
$V_{a}^{j}$ is the co-moving volume available for a source $j$ in a
bin with sizes $\Delta M_{1450}$ and $\Delta z$. Due to the small
number of quasars and small range in redshift, we only use one
redshift bin. In the following section we will discuss the likely
amount of evolution across this bin. The available volume takes into
account the selection functions, $p(M_{1450},z)$, described in Sec.\,\ref{surveys} and
can be calculated as
\begin{displaymath}
V_{a}^{j} ~= \int \hspace{-0.15cm} \int p(M_{1450},z)~ \frac{dV}{dz} ~dz ~dM_{1450}.
\end{displaymath}
The volume element $dV/dz$ accounts for the sky area of the survey(s) at this location in $M_{1450},z$ space. The luminosity function, $\Phi (M_{1450})$, is then calculated as 
\begin{displaymath}
\Phi (M_{1450})~ = \sum^{N}_{j=1} ~\frac{1}{V_{a}^{j}}~ (\Delta M_{1450})^{-1}.
\end{displaymath}

The absolute magnitude bins were chosen to avoid human-induced bias as
much as possible. No attempt was made to get approximately equal
numbers of objects in each bin, since this can bias the results for
small samples.  

Unlike in the work of Jiang et al. (2009) which treated their $z'<21$
and $z'>21$ samples separately , we combine the two SDSS deep stripe
samples of Jiang et al. (2008; 2009) so that the SDSS deep quasars are
in bins determined by their $M_{1450}$ rather than their apparent $z'$
magnitude.

\begin{figure}
\resizebox{0.50\textwidth}{!}{\includegraphics{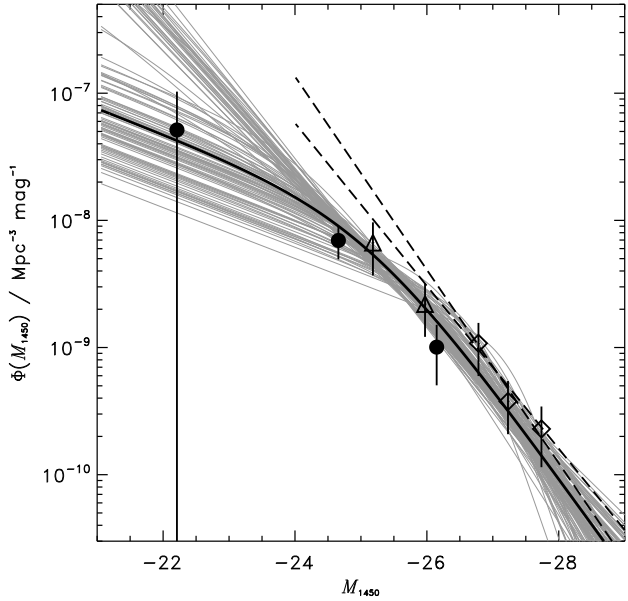}}
\caption{Binned (open diamonds SDSS main, open triangles SDSS deep
  stripe and filled circles CFHQS) and parametric best fit (thick
  line) $z=6$ luminosity functions. The thin grey lines show the
  result of 100 bootstrap resamples. The two dashed lines are the
  power law fits of Jiang et al. (2009).
\label{fig:lf}
}
\end{figure}

Fig.\,\ref{fig:lf} shows the binned luminosity function derived from
the CFHQS and SDSS samples. Data points are plotted at the mean
absolute magnitude of objects within the bin. Separate points are
plotted for the CFHQS, SDSS main and SDSS deep stripe samples, even
though there is some overlap in absolute magnitude of the three
samples. Note that due to a different choice of bins sizes, different
SDSS main sample size (Jiang et al. included some unpublished quasars)
and different cosmological parameters, the SDSS points are not in
exactly the same locations as in Jiang et al. (2009).

The dashed lines in Fig.\,\ref{fig:lf} are the two power law fits to
the binned luminosity function by Jiang et al. (2009), corrected for
our cosmology. The steeper slope line ($\beta=-2.9$) is a fit to the
SDSS main and SDSS deep $z'<21$ binned points and the flatter slope
($\beta=-2.6$) is a fit to the SDSS main and all SDSS deep binned
points in Jiang et al. (2009). These power laws from Jiang et al. lie
above the SDSS binned points we have calculated. The likely reasons
for this difference are because we used different bins and have a
different SDSS main sample size. We set the bins to cover absolute
magnitude ranges without considering the distribution of data points
within each bin. Jiang et al. (2008, 2009) set the edges of their
absolute magnitude bins to be equal to the maximum and minimum
absolute magnitudes of the objects within the bins (L.\, Jiang,
priv. comm.). For very few objects in the bin this necessarily biases
the estimate of the volume available in the bin to low values and
hence unreasonably high space densities. This is particularly apparent
for the brightest SDSS deep stripe point and faintest SDSS main sample
point. A power law fit to our SDSS bins would show a somewhat flatter
slope and lower normalisation than either of those found by Jiang et
al. However, the discussion above highlights the perils of determining
the luminosity function from binned data and we therefore leave any
parametric derivation to the following section.

There are three CFHQS binned points on Fig.\,\ref{fig:lf}. The
faintest point at $M_{1450}\approx -22$ contains the single quasar
from the Deep/SXDS. The other two points contain all the
quasars from the RCS-2/VW. The point at $M_{1450}\approx -24.7$
contains 12 quasars and has by far the smallest statistical error of
any points on the plot. The two RCS-2/VW points lie below the
luminosity function defined by the SDSS main and deep stripe binned
data. However, these data are all consistent within the size of the
1\,$\sigma$ error bars, so we do not consider the SDSS and CFHQS data
to be in conflict. Considering all the data points together, there
appears to be some evidence for a flattening of the luminosity
function going from high to low luminosity. The main evidence for a
break at $M_{1450}\approx -25$ is based on the two faintest CFHQS
bins. Unfortunately, the small sky area probed by the Deep/SXDS and
the single quasar found means that the space density at very low
luminosities is not strongly constrained. We will revisit this issue
in the following section.

\subsection{Parametric luminosity function}

To overcome the limitations of the binned method, we also fitted a
parametric luminosity function model to the data. As shown by the
binned data in Fig.\,\ref{fig:lf} the luminosity function is fit well
by a power law at the high luminosity end, with the possibility of a
break at low luminosity. In keeping with previous work, we therefore
consider the quasar luminosity function at $z=6$ to follow a double
power law function:
\begin{displaymath}
\Phi (M_{1450},z) ~ = \frac{10^{k(z-6)}~  \Phi(M_{1450}^{*})}{10^{0.4(\alpha+1)(M_{1450}-M_{1450}^*)} + 10^{0.4(\beta+1)(M_{1450}-M_{1450}^*)}}.
\end{displaymath}
We include an evolution term because the quasar space density is well
known to decline with redshift from a peak at $z\approx 2.5$
(e.g. Richards et al. 2006). Our sample has only a small redshift
range so we do not attempt to fit for the evolution parameter $k$, but
adopt the value of $k=-0.47$ which is derived from the evolution from
$z=3$ to $z=6$ for the bright end of the luminosity function (Fan et
al. 2001). This rate of evolution corresponds to a factor of 2 decrease in
space density across our redshift range from $z=5.8$ to $z=6.4$. Hence
it is important to include because redshift and absolute magnitude can
be correlated in these color- and magnitude-limited samples
(e.g. Fig.\,\ref{fig:comparrays}).

This leaves four parameters to be determined:the normalisation
$\Phi(M_{1450}^{*})$, the break magnitude $M_{1450}^{*}$, the bright
end slope $\beta$ and the faint end slope $\alpha$. As can be seen in
Fig.\,\ref{fig:lf}, there are very few quasars in our sample at
magnitudes well below the break in the luminosity function. Therefore
we have decided to fix the faint end slope to $\alpha=-1.5$ as
indicated by AGN surveys at lower redshift (Croom et al. 2004; 2009; Hunt et
al. 2004). The remaining three parameters are determined via a maximum
likelihood fit (Marshall et al. 1983). The aim is to minimise the function S which equals $-2 \ln \mathcal{L}$, where $\mathcal{L}$ is the likelihood:
\begin{eqnarray}
\label{eq:qlike}
\lefteqn{\nonumber S ~=~ -2 \sum^{N}_{i=1} ~\ln~ [\Phi(M_{1450\,i},z_{i})~ p(M_{1450\,i},z_{i})]}\\
\nonumber & & +2 \int \hspace{-0.15cm} \int \Phi(M_{1450},z)~ p(M_{1450},z)~ \frac{dV}{dz}~ dz~dM_{1450}
\end{eqnarray}

where the first term is a sum over each quasar and the second is
integrated over the full possible range of redshift and absolute
magnitude. Initial starting values of the free parameters were
estimated from the binned luminosity function. The parameters were
optimised using the amoeba routine (Press et al. 1992). 

\begin{figure}
\hspace{-0.5cm}
\resizebox{0.50\textwidth}{!}{\includegraphics{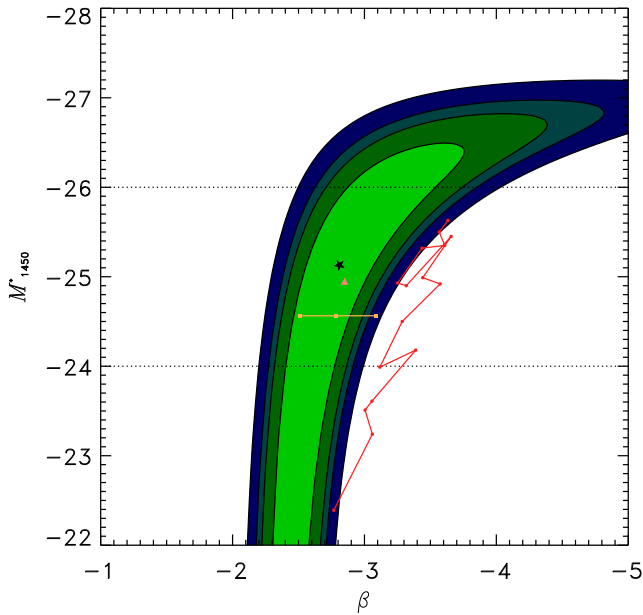}}
\caption{The contours show the confidence regions (68.3\%, 90\%,95.4\%, 99\%) for the bright end slope, $\beta$, and the break absolute magnitude, $M_{1450}^{*}$ at $z=6$. The best-fit $z=6$ parameters are marked with a star. The pink, red and orange symbols show data at lower redshifts. The pink triangle is $z=3.2$ from Siana et al. (2008) based on SWIRE+SDSS. The orange squares joined by a line are from the evolving $\beta$ model fit to the SDSS by Richards et al. (2006) from $z=2.5$ (right) to $z=4.5$ (left). The red dots joined by lines are narrow redshift slices from $z=0.75$ to $z=2.25$ based on the 2SLAQ+SDSS work of Croom et al. (2009).
\label{fig:betamstar}
}
\end{figure}

The best-fit luminosity function parameters are $\Phi(M_{1450}^{*})=
1.14\times 10^{-8}\,{\rm Mpc}^{-3}\,{\rm mag}^{-1}$,
$M_{1450}^{*}=-25.13$, $\beta=-2.81$. This function is plotted on
Fig.\,\ref{fig:lf} as the thick line. It passes through the
$1\,\sigma$ error bars on each of the binned data points showing good
agreement between the maximum likelihood and binned methods. The $z=6$
quasar space density at $M_{1450}=-25$ is a factor of two smaller than
that of the $\beta=-2.6$ fit to binned data of Jiang et
al. (2009). This has important consequences for the quasar ionizing
photon output at $z=6$ as we shall see later.

Uncertainties on the luminosity function parameters were determined
assuming a $\chi^{2}$ distribution of $\Delta S$ ($\equiv S-S_{\mathrm
  min}$) (Lampton, Margon \& Bowyer 1976). Due to the fact that the
parameters are highly correlated the most useful constraints to
consider are those on the joint probability of break absolute
magnitude and bright end slope. Note that the faint end slope was
always fixed at $-1.5$, so any extra uncertainty due to this is
unaccounted for.

Fig.\,\ref{fig:betamstar} plots the $\chi^{2}$ confidence regions for
the combination of break absolute magnitude and bright end slope. The
parameters are obviously correlated such that bright break magnitudes
are consistent with steep bright end slopes. The correlation is easy
to understand because a bright break combined with a steep bright end
slope would give a similar ratio of bright SDSS to faint CFHQS quasars
as a luminosity function with a faint break and flatter bright end
slope. This correlation tends to disappear when the break magnitude is
fainter than $M_{1450}^{*} \approx -25$ because we have few quasars at
such faint levels. The 68.3\% confidence region extends off the bottom
of this plot, which effectively corresponds to an unbroken power law
down to our faintest quasar ($M_{1450}=-22.2$).

The reliability of the $\chi^{2}$ parameter uncertainties were checked
by bootstrap resampling of the data and re-running the maximum
likelihood fit on these samples (Press et al. 1992). 100 bootstrap
trials were performed. A comparison of the fraction of bootstrap
trials within various regions of the confidence contours in
Fig.\,\ref{fig:betamstar} showed very good agreement. 25\% of the
bootstrap trials gave $M_{1450}^{*}>-22$ with $\beta$ in the range
$-2.67$ to $-2.27$, which effectively means that the power law is
unbroken down to our faintest quasar. Although formally allowed by the
data, as mentioned above, this would be quite unusual given the
double-power law form and break magnitudes observed at lower
redshifts. The 100 bootstrap trials are shown in thin grey lines on
Fig.\,\ref{fig:lf} to illustrate the range of possible luminosity
functions allowed by the data. The space density is very well
constrained over the range $-27.5<M_{1450}<-24.7$ with the 1\,$\sigma$
uncertainty being less than 0.1 dex.

In order to try to break the degeneracy between break magnitude and
bright end slope, and to assess evolution of the quasar luminosity
function, we consider the results from studies at lower
redshift. Croom et al. (2009) presented the quasar luminosity function
at $0.4<z<2.6$ based on the 2SLAQ and SDSS surveys. They had
sufficient spread in luminosity and redshift to be able to determine
the luminosity function in narrow ($\Delta z=0.1$) redshift
slices. They found evolution in the break magnitude and steep end
slope (note that $\alpha$ and $\beta$ are swapped in Croom et
al. compared to our definition). The evolution in $\beta$ is quite
mild with a steepening towards higher redshift. Most of the evolution
is in the break magnitude which brightens by 4 magnitudes from $z=0$
to $z=2$ (this is the so-called pure luminosity evolution that
characterised early derivations of the evolving quasar luminosity
function, e.g. Mathez 1976). The break magnitude and $\beta$ found by
Croom et al. (2009) are plotted on Fig.\,\ref{fig:betamstar} (we
converted from $M_{g}(z=2)$ to $M_{1450}$ using the relation in Croom
et al. and adapted for our cosmology). The parameters at $0.75<z<2.25$
are always located at regions in the plot outside of the $z=6$ 95\%
confidence region, with the low redshift slope being steeper than that
at $z=6$ for a given break magnitude.

Richards et al. (2006) determined the quasar luminosity function up to
$z=5$ using the SDSS DR3. At $z>2.4$ they fit a model with a
redshift-dependent bright end slope and fixed break magnitude (due to
the fact they did not have any quasars below the break at high
redshift). Their fit showed the bright end slope flattening from
$\beta=-3.1$ at $z=2.4$ to $\beta=-2.5$ at $z=4.5$. This is shown on
Fig.\,\ref{fig:betamstar} by the orange line with points at $z=2.5,
3.5, 4.5$. Support for the choice of break magnitude adopted by
Richards et al. (2006) comes from the luminosity function derived by
Siana et al. (2008) using SDSS and SWIRE. In fact, the break magnitude
and slope found by Siana et al. (2008) for $z=3.2$ (pink triangle on
Fig.\,\ref{fig:betamstar}) are very close to the minimum $\chi^{2}$
best fit at $z=6$. This would suggest little evolution in these
parameters from $z=3$ to $z=6$, however, there is a
considerable range at 95\% confidence for $z=6$. 

The studies at lower redshift have shown that at all redshifts in the
range $1<z<4$, the break magnitude lies between $-26<M_{1450}<-24$.
Therefore we assume that this also holds true at $z=6$ in order to
provide tighter constraints on $\beta$. Under this assumption we find
the 95\% confidence ranges for $\beta$ to be $-2.9 < \beta < -2.3$ for
$M_{1450}^{*}=-24$ and $-3.8 < \beta < -2.7$ at $M_{1450}^{*}=-26$.

Throughout all the above analysis, the faint end slope was fixed at
$\alpha=-1.5$ based on evidence for $\alpha$ at lower redshift (Croom
et al. 2004; 2009; Hunt et al. 2004). We have checked to see how our
results would change if instead we adopted a different value for
$\alpha$. We re-ran the maximum likelihood fitting assuming
$\alpha=-1.8$ and determined best-fit parameters of
$\Phi(M_{1450}^{*})= 2.55\times 10^{-9}\,{\rm Mpc}^{-3}\,{\rm
  mag}^{-1}$, $M_{1450}^{*}=-26.39$ and $\beta=-3.26$. Although these
best-fit parameters appear substantially different to those for
$\alpha=-1.5$, the two luminosity functions are very similar over the
luminosity range covered by the SDSS and CFHQS quasars. The steeper
bright end slope and higher break luminosity are shifted along the
same direction as the correlation in Fig.\,\ref{fig:betamstar}. The
best-fit break at $M_{1450}^{*}=-26.39$ is more luminous than as
measured at lower redshifts and as noted above we find it unlikely
that the break would be this luminous at $z=6$. Future large area,
deep $z \sim 6$ quasar surveys are necessary to determine the faint
end slope.

\subsection{Consistency of luminosity function with redshift distributions}

These colour-selected quasar samples can only find quasars within a
certain redshift range as shown in Figs.\,\ref{fig:photcomp} and
\ref{fig:comparrays}. The successful derivation of a luminosity
function from these data depends upon accurately determining how the
completeness depends upon redshift and luminosity. In order to check
the consistency of the observed redshift distributions with the
selection function we have used the selection functions and the
best-fit luminosity function to generate expected redshift
distributions for 3 samples: SDSS main, SDSS deep stripe (combined
$z'<21$ and $z'>21$) and CFHQS RCS-2/VW. These expected redshift
distributions were then compared to the observed redshift
distributions. A Kolmogorov-Smirnov test was applied to determine the
probability that the redshift distributions were consistent. The
results showed that all samples are consistent with their expected
redshift distributions (at probabilities 0.48, 0.78, 0.27,
respectively). Jiang et al. (2009) noted that their SDSS deep $z'>21$
sample had no quasars at $z>6.1$ and showed that this lack of
high-redshift quasars was not statistically significant. Our findings
agree and show that all the samples have redshift distributions as expected from
the colour selection criteria.

\section{Applications of the luminosity function}

\subsection{The quasar intergalactic ionizing flux at $z=6$}

The constraints placed on the low luminosity end of the $z=6$ quasar
luminosity function here are considerably tighter than in previous
works (Willott et al. 2005a; Shankar \& Mathur 2007; Jiang et
al. 2009). Therefore it is useful to consider the ionizing photon
output of the quasar population at $z=6$. We use the 100 bootstrap
trial luminosity functions to determine the plausible statistical
range of the photon output. Since the luminosity function is not
constrained at $M_{1450}^{*}>-22$, we set all the bootstrap models
with break magnitudes fainter than this to have a break to
$\alpha=-1.5$ at $M_{1450}^{*}=-22$. Note however that we do not
include ranges on other important parameters, such as the faint end
slope $\alpha$ (assumed -1.5 here), the quasar UV spectral slope
(assumed that of Telfer et al. 2002) or the photon escape fraction
(assumed 100\%).

\begin{figure}
\resizebox{0.48\textwidth}{!}{\includegraphics{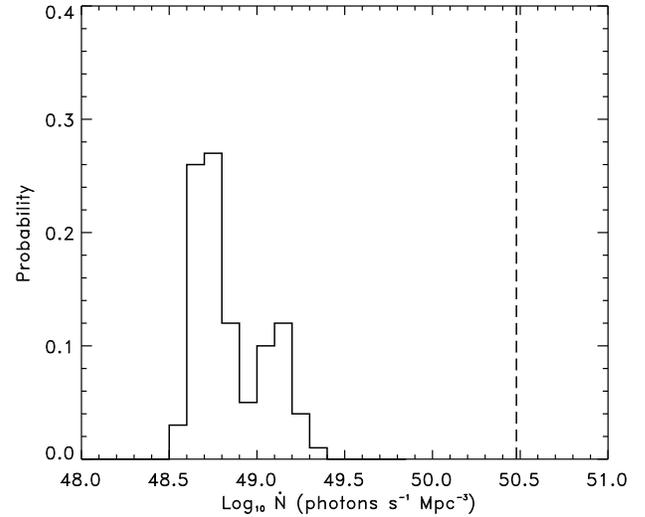}}
\caption{The probability distribution for the ionizing photon emission
  rate density from quasars at $z=6$. The distribution comes from the
  range of possible luminosity functions consistent with the data
  determined by the bootstrap method. The dashed line shows the
  required photon emission rate density to balance combinations and
  keep the universe ionized.
\label{fig:ion}
}
\end{figure}

The resulting probability distribution in ionizing photon rate density
is shown in Fig.\,\ref{fig:ion}. The peak of the distribution is at
$4\times10^{48}$\,photons\,s$^{-1}$\,Mpc$^{-3}$ and the median is at
$5\times10^{48}$\,photons\,s$^{-1}$\,Mpc$^{-3}$. The probability
distribution is bimodal because the group of bootstrap luminosity
functions with $M_{1450}^{*}>-22$ have a large number of faint quasars
and total emission rate density of
$>10^{49}$\,photons\,s$^{-1}$\,Mpc$^{-3}$. Our results are
significantly lower than that previously assumed of
$2\times10^{49}$\,photons\,s$^{-1}$\,Mpc$^{-3}$ (Meiksin 2005; Bolton
\& Haehnelt 2007). Because low-luminosity quasars could dominate the
ionizing flux, the bootstrap resampling was also run for a steeper
faint end slope of $\alpha=-1.8$ and in this case the probability
distribution is only shifted 0.1 dex higher than for $\alpha=-1.5$.

Also, plotted on Fig.\,\ref{fig:ion} is a dashed line to denote the
photon rate density required in order to balance recombinations and
keep the universe ionized. We follow Meiksin (2005) to calculate this
number and assume that the clumpiness factor $C=5$. It is evident from
this plot that the quasar population, even including the possibility
of a large number of faint quasars, is insufficient to get even close
to the required photon emission rate density. Our estimate of the
photon rate density is between 20 and 100 times lower than the
required rate.  This is consistant with the constraints placed on this
rate by the unresolved X-ray background (Djikstra et al. 2004). In
comparison, the known galaxy luminosity function at $z=6$ shows that
galaxies provide between $10^{50}$ and
$4\times10^{50}$\,photons\,s$^{-1}$\,Mpc$^{-3}$ (Ouchi et al. 2009).

\begin{figure*}
\resizebox{0.98\textwidth}{!}{\includegraphics{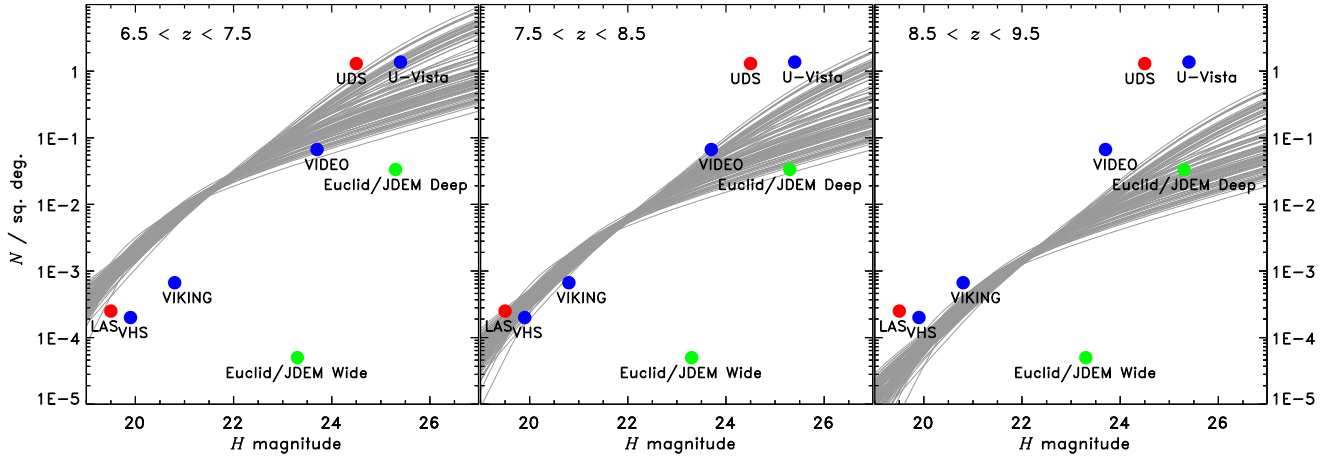}}
\caption{Predictions for $z>6.5$ quasar surveys based on our
  luminosity function. Each plot shows, for a particular redshift
  slice, the sky surface density of quasars brighter than the $H$
  magnitude limit. The grey lines show the 100 bootstrapped luminosity
  functions consistent with our data evolved to higher redshift as
  described in the text. The symbols show the depths of near-IR
  surveys and the sky density necessary for them to detect one
  quasar in the redshift slice (UKIDSS -- red, VISTA -- blue, {\it Euclid}/{\it JDEM} -- green).
\label{fig:predict}
}
\end{figure*}

\subsection{Search strategies for quasars at even higher redshifts}

Now that $\sim 50$ quasars are known at redshifts $5.7<z<6.5$, the
next challenge is to identify quasars at even higher redshifts. This
is particularly important given the possible rapid evolution in the
ionization state of the IGM at these redshifts (e.g. Fan et al
2006c). As has been well documented, it is challenging to discover
quasars at higher redshifts as the \lya\ line and continuum move into
the near-IR. Several large surveys are now underway or planned aiming
to discover quasars at $z>7$. Given our latest determination of the
luminosity function at $z=6$, we can predict the number of quasars
that may be detectable at higher redshifts. The main aim of these
predictions is to investigate the optimum observing strategy in a
fixed observing time, i.e. wide--shallow vs narrow--deep.

We calculate the density of quasars on the sky as a function of
limiting apparent $H$ magnitude in three redshift slices ($6.5<z<7.5$,
$7.5<z<8.5$, $8.5<z<9.5$). The calculations are performed for all 100
of the bootstrap resampled luminosity functions. The evolution is
assumed to continue as density evolution at the same exponential rate
as determined for luminous quasars by Fan et al (2001) from $z=3$ to
$z=6$ (i.e. $10^{k(z-6)}$ with $k=-0.47$). The actual quasar evolution
is obviously completely unknown and it is possible that some
luminosity-dependence of the evolution occurs over this redshift
range.  Theoretical models based on Eddington-limited accretion of the
$z=6$ quasar population traced back to earlier epochs predict that the
black hole accretion rate density is somewhere between 10 and 100
times lower at $z=9$ than $z=6$ (Li et al. 2007; Sijacki et
al. 2009). Note that these models are focussed on overdense regions,
due to simulation limitations, so may not represent typical
regions. However, their evolution compares well with our factor of
$\sim 30$ evolution in space density from $z=9$ to $z=6$. 

Recent deep imaging from WFC3 onboard \hst and from the ground has
enabled the identification of galaxies up to $z=8.5$ (Ouchi et
al. 2009; Bouwens et al. 2009; Oesch et al. 2009). Assuming these
galaxies are not mostly low redshift interlopers (see Capak et
al. 2009), the evolution in the galaxy luminosity function from $z=6$
to $z=7$ and $z=8$ is not too rapid. Bouwens et al. (2009) and McLure
et al. (2009) show the faint end evolution from $z=6$ to $z=8$ is less
than a factor of ten. Furthermore, Labbe et al. (2009) show the
stellar mass density evolves only by a factor of three between $z=6$
and $z=7$. Although it is unclear how the evolution of mostly low
luminosity galaxies relates to that of luminous quasars, these results
are consistent with the evolution for quasars we have adopted.

Fig.\,\ref{fig:predict} shows the 100 bootstrap quasar count
predictions in each of the three redshift slices. The expected counts
are most strongly constrained for quasars with similar absolute
magnitude to the SDSS and CFHQS quasars, corresponding to observed
magnitudes 20 to 23. Fontanot et al. (2007a) made similar predictions
for the near-IR quasar number counts based on extrapolation of the
luminosity functions of Fontanot et al. (2007b) at $z \sim 4$ and
Shankar \& Mathur (2007) at $z\sim 6$. They did not show results for
the bright end of the luminosity function at apparent magnitudes
$H<24$. At $H>24$ our results are consistent with the predictions
based on the evolving model of Fontanot et al. (2007b).

We now consider the specifics of various ongoing and future near-IR
sky surveys which might be able to discover high-$z$ quasars.  All
limiting magnitudes are given as $10\sigma$ AB magnitudes, because
experience of the CFHQS shows that this level of photometric accuracy
in the bands redward of the \lya\ line is necessary for reliable
colour selection. It should also be noted that we simply consider
whether these surveys cover sufficient depth and area at $H$ band to
be able to detect the quasars. Successful quasar discovery
depends upon colour criteria that isolate the quasars from all
contaminants such as brown dwarfs. This usually means that data in
filters shortward of the \lya\ break must go 1 to 2 magnitudes deeper
than the $H$ band data.

The UKIRT Infrared Deep Sky Survey (UKIDSS; Lawrence et al. 2007)
began in 2005 and should be completed in a few years. It has several
components of which the two we consider here are the Large Area Survey
(LAS; 4000 sq. deg. to $H=19.5$) and the Ultra Deep Survey (UDS; 0.77
sq. deg. to $H=24.5$). A primary goal of the LAS is to find $z=7$
quasars (Hewett et al. 2006) and it so far has found two at $z \approx
6$ (Venemans et al. 2007; Mortlock et al. 2009). In
Fig.\,\ref{fig:predict}, these surveys are plotted alongside the 100
bootstrap predictions such that the location on the plot indicates
that one quasar would be found in the survey in that redshift
slice. At $z\sim 7$ the LAS falls below the predictions indicating
that between 1 and 3 quasars would be expected in this survey at this
redshift. The UDS lies right at the upper edge of the predicted quasar
counts meaning that it is likely that it will contain $\ltsimeq 1$
quasar at $6.5<z<7.5$. Note that the UDS forms part of the deep
component of the CFHQS and the lone deep CFHQS quasar at $z=6.01$
actually lies within the UDS. At higher redshifts, the LAS may contain
$\sim 1$ quasar at $7.5<z<8.5$, but it is very unlikely for the UDS to
contain any higher redsift quasars. In conclusion, the wide area
component of UKIDSS is much better suited to discovering high-$z$
quasars than the deep component.

Set to commence imminently, the ESO VISTA Telescope surveys (Sutherland
2009) are more efficient than UKIDSS due to the use of a larger camera. There are
four components of the VISTA surveys that could potentially detect
high-$z$ quasars. These are VISTA Hemisphere Survey (VHS; 5000
sq. deg. to $H=19.9$ coincident with the Dark Energy Survey in the
optical), VISTA Kilo-Degree Infrared Galaxy Survey (VIKING; 1500
sq. deg. to $H=19.9$ coincident with VST KIDS in the optical), VISTA
Deep Extragalactic Observations Survey (VIDEO; 15 sq. deg. to
$H=23.7$) and UltraVISTA (0.73 sq. deg. to $H=25.4$ in the COSMOS
field). As shown on Fig.\,\ref{fig:predict} these surveys should be
deep enough at $H$ to detect many quasars at $6.5<z<7.5$ and a few at
$7.5<z<8.5$. The most suitable surveys are the wider area surveys of
VHS and VIKING, rather than VIDEO and UltraVISTA. This is a
consequence of the relatively shallow $z=6$ luminosity function we
have found. With a slope of $\beta>-3.0$, more quasars can be found in
a wide area shallow survey than a narrow area deep survey in a fixed
observing time. However, we caution that if the bright end of the
luminosity function steepens at high redshift, due for example to the
lack of time available for the build-up of the required mass black
holes, then this would reverse the situation. Therefore VIKING is
probably more likely to yield $z \sim 8$ quasars than VHS. Finally, we
note that none of the VISTA surveys are likely to detect $z\sim 9$
quasars.

What about the longer-term future? Several ambitious dark energy
projects are being planned such as {\it Euclid} and {\it JDEM}. These would use
some combination of supernovae, baryonic acoustics oscillations and
weak lensing to determine the parameters governing the evolution of
dark energy. In order to make these measurements it is necessary to
make deep surveys of large sky areas in the near-IR and therefore such
surveys would also be useful to detect the most distant quasars. Given
the early planning stage it is not known yet how much sky area will be
covered to what depth. For the purposes of this exercise we use a
preliminary {\it Euclid} plan which include a wide 20\,000 sq. deg. imaging
survey to $H=23.3$ for weak lensing and a deep imaging survey of 30
sq. deg. to $H=25.3$ for supernovae.  The green circles on
Fig.\,\ref{fig:predict} show that the wide survey is much more
effective at detecting distant quasars due to the flatter faint end
slope (note this is not strictly a fair comparison because to these
depths and areas in a single filter, the wide survey requires 17 times
as long as the deep survey). The wide survey would be able to detect
$\sim 1000$ quasars at $z\sim 7$, $\sim 400$ at $z\sim 8$ and $\sim
100$ at $z \sim 9$. As before, we caution that sufficiently deep data
at filters below the \lya\ break are necessary to provide accurate
colours, so this is an upper limit to the number of quasars that might
be discoverable. We do not consider even deeper surveys over small sky
areas with {\it JWST}, because the $z=6$ luminosity function is not
well constrained for such low luminosity quasars.

\section{Conclusions}

We have presented discovery data for nine new quasars in the CFHQS,
bringing the total number so far to 19. The CFHQS is nearing
completion and ongoing follow-up may find a few more CFHQS quasars.
Further observations of CFHQS quasars are being pursued to understand reionization,
black hole mass growth and host galaxy evolution.

The CFHQS and SDSS surveys were combined to derive the $z=6$ quasar
luminosity function from a sample of 40 quasars. The normalisation of
the luminosity function is found to be $\approx 40$\% lower than that
determined previously using binned data (Jiang et al. 2008; 2009). The
luminosity function is well constrained down to $M_{\rm 1450} \approx
-24$ but has a large uncertainty at $M_{\rm 1450} \approx -22$ where
our faintest quasar is. Future larger sky area, deep surveys, such as
VISTA VIDEO, are necessary to constrain this part of the $z=6$ quasar
luminosity function. We are also working on measuring black hole
masses for $z>6$ quasars. In a future paper we will combine our
knowledge of the luminosity function and black hole mass distribution
to consider the growth of black holes at this early epoch.

We have calculated the quasar ionizing flux based on possible
realizations of the $z=6$ quasar luminosity function and found that
the ionizing photon output of quasars is between 20 and 100 times
lower than the rate required to keep the universe ionized. Predictions
have been made for the number of $z>6.5$ quasars that might be found
by current and future surveys. Whilst UKIDSS may find a few quasars at
$z\sim 7$, the VISTA surveys should find many more and even some up to
$z\sim 8$. Obtaining large samples of $z\sim 8$ quasars and pushing to
even higher redshifts will require even more ambitious projects such
as {\it Euclid} or {\it JDEM}.

\acknowledgments

Based on observations obtained with MegaPrime/MegaCam, a joint project
of CFHT and CEA/DAPNIA, at the Canada-France-Hawaii Telescope (CFHT)
which is operated by the National Research Council (NRC) of Canada,
the Institut National des Sciences de l'Univers of the Centre National
de la Recherche Scientifique (CNRS) of France, and the University of
Hawaii. This work is based in part on data products produced at
TERAPIX and the Canadian Astronomy Data Centre as part of the
Canada-France-Hawaii Telescope Legacy Survey, a collaborative project
of NRC and CNRS. Based on observations obtained at the Gemini
Observatory, which is operated by the Association of Universities for
Research in Astronomy, Inc., under a cooperative agreement with the
NSF on behalf of the Gemini partnership: the National Science
Foundation (United States), the Particle Physics and Astronomy
Research Council (United Kingdom), the National Research Council
(Canada), CONICYT (Chile), the Australian Research Council
(Australia), CNPq (Brazil) and CONICET (Argentina). This paper uses
data from Gemini programs GS-2008B-Q-23, GS-2009A-Q-3, GS-2009B-Q-25,
GN-2008B-Q-42 and GN-2009A-Q-2. Some of the data presented herein were
obtained at the W.M. Keck Observatory, which is operated as a
scientific partnership among the California Institute of Technology,
the University of California and the National Aeronautics and Space
Administration. The Observatory was made possible by the generous
financial support of the W.M. Keck Foundation. Based on observations
made with the ESO New Technology Telescope at the La Silla
Observatory. Based on observations made with the Nordic Optical
Telescope, operated on the island of La Palma jointly by Denmark,
Finland, Iceland, Norway, and Sweden, in the Spanish Observatorio del
Roque de los Muchachos of the Instituto de Astrofisica de Canarias.
Thanks to Howard Yee and the rest of the RCS2 team for sharing their
data and to the queue observers at CFHT and Gemini who obtained data
for this project. Thanks to Xiaohui Fan, Linhua Jiang and Scott Croom
for providing information relating to their work on the luminosity
function and Jasper Wall for interesting discussions. Thanks to the
anonymous referee for suggestions which improved this manuscript.


\clearpage

\appendix

\section{Finding charts}

Fig.\,\ref{fig:finders} presents $3' \times 3'$ finding charts for the
CFHQS quasars. All images are centered on the quasars and have the
same orientation on the sky. These are the CFHT MegaCam $z'$-band
images in which the quasars were first identified. MegaCam has gaps
between the CCDs and these data were not dithered so the gaps remain
and are evident in 5 images where the quasars lie close to the edge of
a CCD. For CFHQS\,J0050+3445, CFHQS\,J0136+0226, CFHQS\,J0316-1340 and
CFHQS\,J1059-0906, the Megacam data are a single exposure leading to
many cosmic rays in the final images.

\begin{figure*}[b]
\hspace{1.1cm}
\resizebox{0.88\textwidth}{!}{\includegraphics{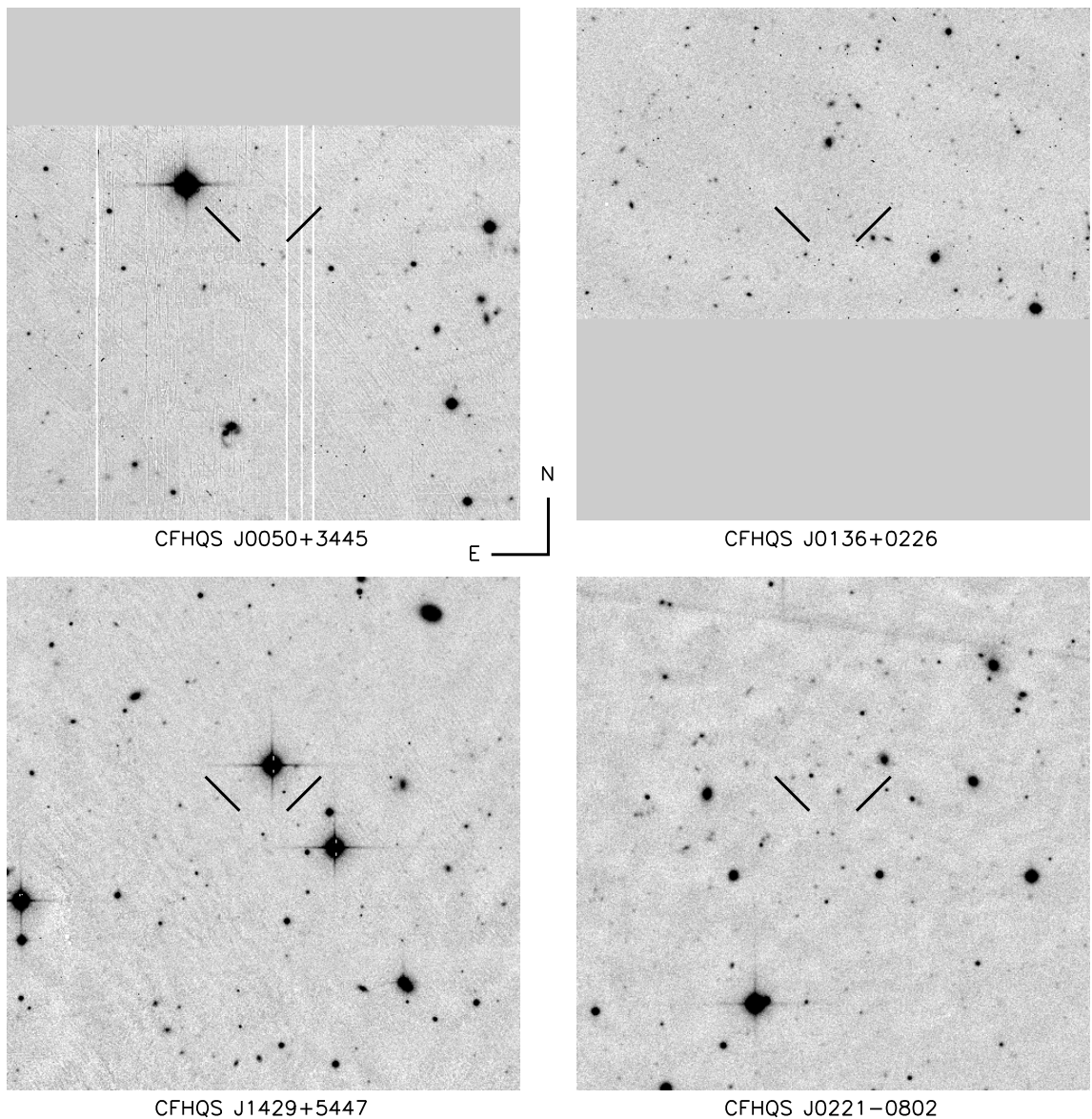}}
\caption{ $z'$-band finding charts for the CFHQS quasars.
\label{fig:finders}
}
\end{figure*}

\addtocounter{figure}{-1}

\begin{figure*}[b]
\hspace{1.1cm}
\resizebox{0.88\textwidth}{!}{\includegraphics{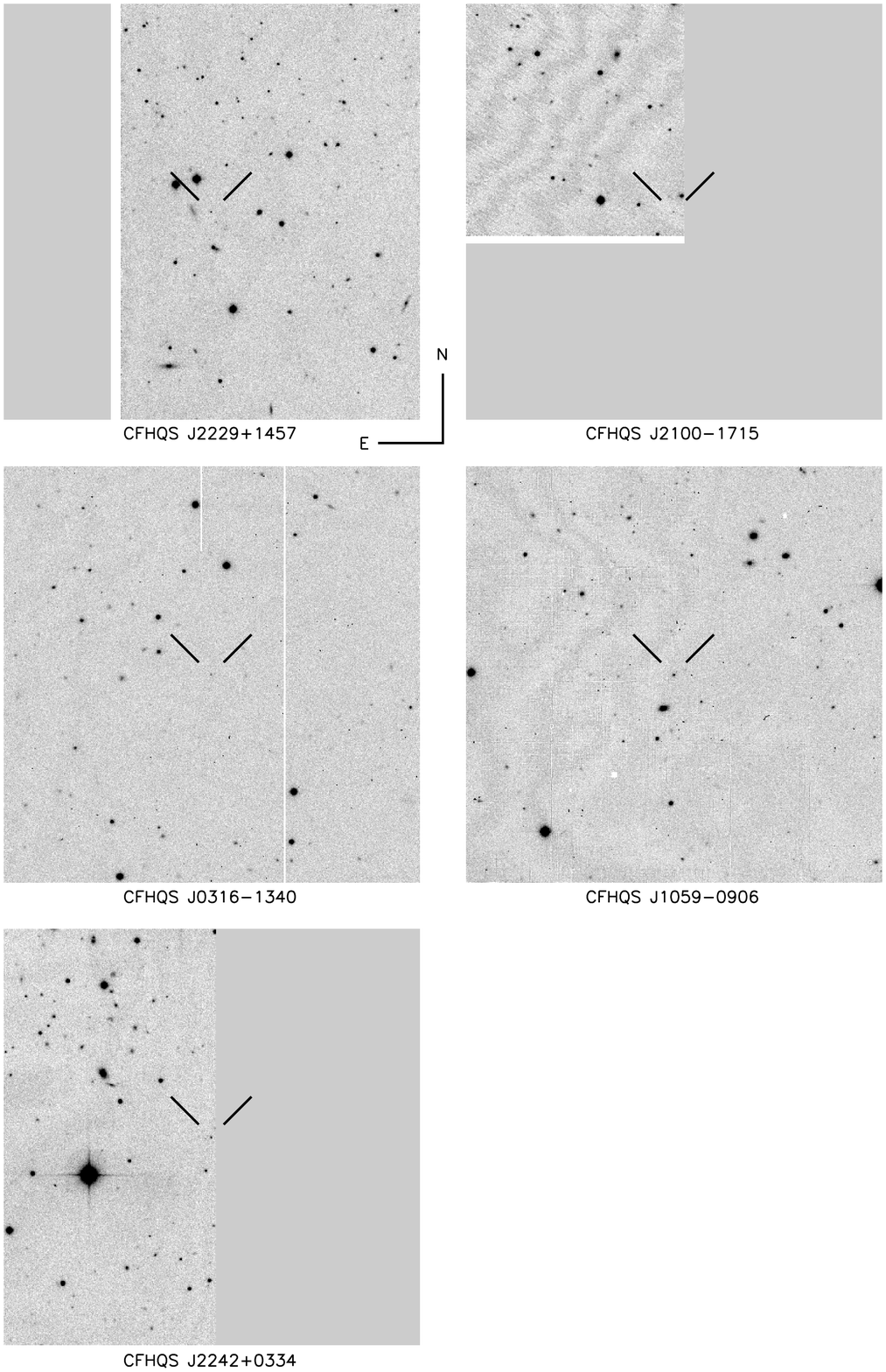}}
\caption{ (cont.)
}
\end{figure*}

\end{document}